\documentclass[fleqn,usenatbib]{mnras}

\usepackage[T1]{fontenc}
\usepackage{ae,aecompl}
\usepackage{graphicx}	
\usepackage{amsmath}	
\usepackage{amssymb}	

\title{On physical interpretations of the reference transit radius of gas-giant exoplanets}

\author[Heng]{
Kevin Heng$^{1}$\thanks{E-mail: kevin.heng@csh.unibe.ch (KH)}
\\
$^{1}$University of Bern, Center for Space and Habitability, Gesellschaftsstrasse 6, CH-3012, Bern, Switzerland}

\date{Accepted 2019 September 27. Received 2019 September 17; in original form 2019 January 21}

\pubyear{2019}

\begin{document}
\label{firstpage}
\pagerange{\pageref{firstpage}--\pageref{lastpage}}
\maketitle

\begin{abstract}
Two theoretical quandaries involving transmission spectra of gas-giant exoplanets are elucidated.  When computing the transit radius as a function of wavelength, one needs to specify a reference transit radius corresponding to a reference pressure.  Mathematically, the reference transit radius is a constant of integration that originates from evaluating an integral for the transit depth.  Physically, its interpretation has been debated in the literature.  \cite{je18} suggested that the optical depth is discontinuous across, and infinite below, the reference transit radius.  \cite{bs17,bs18} interpreted the spherical surface located at the reference transit radius to represent the boundary associated with an opaque cloud deck.  It is demonstrated that continuous functions for the optical depth may be found.  The optical depth below and at the reference transit radius need not take on special or divergent values.  In the limit of a spatially uniform grey cloud with constant opacity, the transit chord with optical depth on the order of unity mimics the presence of a ``cloud top".  While the surface located at the reference pressure may mimic the presence of grey clouds, it is more natural to include the effects of these clouds as part of the opacity function because the cloud opacity may be computed from first principles.  It is unclear how this mimicry extends to non-grey clouds comprising small particles.
\end{abstract}


\section{Introduction}

An exoplanet transiting its star produces an obscuring disc corresponding to some transit radius, which is generally a function of wavelength.  This transit radius corresponds to a sightline from the observer to the star that is a chord (in the mathematical sense), with an optical depth on the order of unity, passing through the exoplanetary atmosphere.  In the limit of an isothermal transit chord and constant acceleration due to gravity, the transit radius is \citep{fortney05,lec08,ds13,heng15,bs17,hk17,je18}
\begin{equation}
R = R_0 + H \left[ \gamma + E_1(\tau_0) + \ln{\tau_0} \right],
\label{eq:main}
\end{equation}
where $R_0$ is a reference transit radius, $H$ is the isothermal pressure scale height, $\gamma$ is the Euler-Mascheroni constant, $\tau_0$ is the optical depth corresponding to the reference transit radius and $E_1(\tau_0)$ is the exponential integral of first order (e.g., \citealt{abram,arfken})
\begin{equation}
E_1(\tau_0) \equiv \int^\infty_1 y^{-1} e^{-y \tau_0} ~dy,
\end{equation}
which has the mathematical property that 
\begin{equation}
\lim_{\tau_0 \rightarrow \infty} E_1(\tau_0) = 0.
\end{equation}

Mathematically, $R_0$ is a constant of integration that results from evaluating an integral.  Its physical interpretation has been debated in the literature.  The goal of the current study is to elucidate two quandaries involving the physical interpretation of $R_0$ and provide possible resolutions to these quandaries.

\subsection{Quandary 1: is the optical depth discontinuous across the reference transit radius?}

The first quandary concerns whether the reference transit radius corresponds to a special physical location within the exoplanet, across which the optical depth is discontinuous.  Consider only gas-giant exoplanets without rocky surfaces such that a discontinuity associated with the interface between the atmosphere and rocky surface cannot be claimed.

\cite{hk17} reasoned that the reference optical depth ($\tau_0$) does not need to take on any particular value, but one may \textit{choose} a value of $R_0$ such that $\tau_0 \gg 1$.  Such a choice implies that the $E_1(\tau_0)$ term in equation (\ref{eq:main}) must vanish.  The obscuring disc has the area \citep{ds13,bs17,hk17},
\begin{equation}
\pi R^2 = \pi R_0^2 + A\left(R_0, \infty \right),
\end{equation}
where $\pi R_0^2$ is the area of the secondary obscuring disc corresponding to the reference transit radius, $A(R_0,\infty)$ is the area of a thin annulus defined by \citep{brown01}
\begin{equation}
A\left(r_1,r_2\right) = \int^{r_2}_{r_1} \left( 1 - e^{-\tau} \right) 2 \pi r ~dr,
\end{equation}
$\tau(r)$ is the optical depth and $r$ is the radial coordinate.

\cite{je18} offered an alternative explanation motivated by equation (3) of \cite{ds13} as a starting point.  They reasoned that, since
\begin{equation}
\pi R^2 = A\left(0,\infty \right) = A\left(0,R_0\right) + A\left(R_0, \infty\right),
\end{equation}
and one necessarily needs to have
\begin{equation}
A\left(0,R_0\right) = \int^{R_0}_0 \left( 1 - e^{-\tau} \right) 2 \pi r ~dr = \pi R_0^2,
\label{eq:main2}
\end{equation}
this implies that the atmosphere immediately below the reference transit radius must possess large optical depths.  Specifically, they remarked that, ``$R_0$ satisfies the condition of being a radius below which the planet is fully opaque."  This quoted statement contains a footnote that states, ``If $R_0$ is not chosen to be at an optically thick region (i.e., a region where $\tau \rightarrow \infty$), then it is not possible to write $A(0,R_0)=\pi R_0^2$."

At face value, it seems challenging to reconcile these two viewpoints.  Equation (\ref{eq:main2}) indeed trivially integrates to yield $\pi R_0^2$ if one allows $\tau \rightarrow \infty$ within the integrand.  However, if one asserts that the optical depth needs to be a continuous function, then this implies that $\tau_0 \gg 1$ and the $E_1(\tau_0)$ term in equation (\ref{eq:main}) is permanently absent.\footnote{The final equation in \cite{je18} is an expression for $\pi R^2$ that contains this $E_1(\tau_0)$ term.}  In order to assert that $\tau \rightarrow \infty$ \textit{and} retain the $E_1(\tau_0)$ term in equation (\ref{eq:main}), one has to assume that the optical depth is a discontinuous, piecewise function, 
\begin{equation}
\tau = 
\begin{cases}
\tau_0 e^{\left(R_0-r\right)/H} & r \ge R_0 \\
\infty & 0 \le r < R_0,
\end{cases}
\label{eq:tau1}
\end{equation}
which one may argue lacks generality.  In the absence of a rocky surface, the origin of this discontinuity is unclear.

The first goal of the present study is to reconcile these viewpoints and demonstrate that the optical depth need not be discontinuous across $R_0$.  In the limit of constant opacity, one may demonstrate that $A(0,R_0) \simeq \pi R_0^2$ for any value of $\tau_0$.

\subsection{Quandary 2: does the reference transit radius correspond to an opaque cloud deck?}

The spherical surface associated with $R_0$ has previously been interpreted by \cite{bs17,bs18} to represent the boundary associated with an opaque (optically thick) cloud deck.  Furthermore, \cite{bs17,bs18} claim that variations of equation (\ref{eq:main}), as derived by \cite{lec08} and \cite{ds13}, are valid only for describing cloudfree atmospheres.  For example, the abstract of \cite{bs17} states, ``Although the formalism of Lecavelier des Etangs et al. is extremely useful to understand what shapes transmission spectra of exoplanets, it does not include the effects of a sharp change in flux with altitude generally associated with surfaces and optically thick clouds."  As another example, Section 2.6 of \cite{bs18} states, ``Until recently, the few analytical formalisms \citep{lec08,ds13} attempting to explain what shapes exoplanet transmission spectra could only do so for clear atmospheres."

When computing the transmission spectrum, one needs to specify the cross section or opacity (cross section per unit mass) as a function of wavelength, temperature and pressure.  Physically, the opacity function includes contributions from the extinction (absorption and scattering) of radiation by atoms, ions, molecules and aerosols/hazes/clouds, whether in the form of spectral lines or a continuum.  These contributions are weighted by their relative abundances (i.e., mass or volume mixing ratios).  Sources of spectral continua include collision-induced absorption and Rayleigh scattering.

The \textit{shape} of the continuum due to extinction by clouds depends on the size of the constituent particles.  A cloud particle is small or large only in comparison to the wavelength of radiation it is absorbing or scattering.  Let the radius of a spherical cloud particle be $r_{\rm cloud}$ and the wavelength be $\lambda$.  When $2 \pi r_{\rm cloud} / \lambda \ll 1$ (small particle), one is in the limit of Rayleigh scattering.  When $2 \pi r_{\rm cloud} / \lambda \gg 1$ (large particle), the opacity is roughly constant and the cloud is ``grey".  These are the principles of Mie theory \citep{mie}, which is more than a century old.  (See, e.g., \citealt{p10} or \citealt{kh18} for modern renditions of it.)

A spatially uniform cloud consisting of large particles may be represented by a constant opacity.  A simple thought experiment will illustrate that, even in this scenario, the cloud naturally produces a boundary that is automatically achieved by radiative transfer.  This is because, at each wavelength, the transmission spectrum picks out the $\tau \sim 1$ transit chord (where $\tau$ is the \textit{chord} optical depth).  Assuming that a radial pressure gradient exists within the atmosphere, the transit chord corresponds to a ``cloud top" pressure of \citep{heng16}
\begin{equation}
P = \frac{0.56 g}{\kappa} \sqrt{\frac{H}{2 \pi R}},
\label{eq:p_chord}
\end{equation}
where $g$ is the acceleration due to gravity and $\kappa$ is the opacity.  Even though the opacity is constant, the cloud is optically thin at lower pressures or higher altitudes and exerts a negligible influence on the spectrum.  It is the same radiative transfer principle for why one observes an edge to the Sun, even though no sharp boundary exists.  This thought experiment suggests that as long as an opacity function may be specified in the formula for the transit radius, the formula may be used to model cloudy atmospheres, contrary to the claim of \cite{bs17,bs18}.

The second goal of the present study is to demonstrate that it is not necessary to impose a boundary associated with an opaque cloud via the reference transit radius, even though it is possible for the surface associated with $R_0$ to \textit{mimic} the effects of a grey cloud deck.  Such mimicry does not straightforwardly extend to non-grey clouds consisting of small particles.

\section{Optical depths from polytropes}

It is useful to visualize the gas-giant exoplanet as consisting of two regions.  The region corresponding to $0 \le r \le R_0$ is referred to as the ``interior" of the exoplanet and it encompasses the vast majority of its mass.  The region corresponding to $r \ge R_0$ is referred to as the ``atmosphere" of the exoplanet and the mass enclosed is negligible.  The demarcation between the two regions is not meant to be sharp.  A different set of approximations is applied to each region.  The ideal gas law is expected to be a good approximation within the atmosphere, but it breaks down deeper into the interior as the pressure increases.

Within the interior of the exoplanet, the solutions to the Lane-Emden equation are used to describe the mass density profile, $\rho(r)$ (Chapter 4 of \citealt{chandra}).  Analytical solutions to the Lane-Emden equation exist for polytropes with indices of 0, 1 and 5.  The current study examines only the first two cases, which correspond to the simplest assumption (constant $\rho$) and a reasonable approximation for hydrogen-helium mixtures at high pressures (e.g., Figure 2 of \citealt{s82}).  Upon specifying $\rho(r)$, one may then evaluate the optical depth,
\begin{equation}
\tau = \int \rho \kappa ~dr.
\end{equation}
Since the opacity function for $0 \le r \le R_0$ cannot be easily specified because of poorly known physics (e.g., \citealt{s82,g05,valencia13}), $\kappa$ is assumed to be constant in this study. 

\subsection{Polytrope of index 0}

A polytrope of index 0 corresponds to a constant mass density, $\rho$ (Chapter 4 of \citealt{chandra}).  While this approximation lacks physical realism, it serves as a mathematical prelude to the more realistic case of a polytrope of index 1.  Furthermore, one may argue that assuming a constant $\rho$ is no worse than assuming a constant optical depth for $0 \le r \le R_0$, i.e., equation (\ref{eq:tau1}).  

By demanding that $\tau = \tau_0$ at $r=R_0$, one obtains
\begin{equation}
\tau = \tau_c \left( 1 - \frac{r}{R_0} \right) + \tau_0,
\label{eq:tau_poly0}
\end{equation}
where $\tau_c  \equiv \rho \kappa R_0$.  At $r=0$, one obtains an optical depth of $\tau_c + \tau_0$.  It is important to emphasise that $\tau_0$ is the ``zero point" for the optical depth, whereas $\tau_c$ is the \textit{difference} in optical depth between the center of the exoplanet and $r=R_0$.  It is analogous to the distinction between displacement and distance.  Thus, we expect $\tau_c$ to be large, but no assumption needs to be made on $\tau_0$.  Demanding that $\tau_c \ggg 1$ is the same as assuming
\begin{equation}
l_{\rm mfp} \lll R_0,
\end{equation}
where $l_{\rm mfp} = 1/\rho \kappa$ is the photon mean free path.  The optical depth is neither discontinuous nor constant, as it goes from a value of $\tau_c$ at the center of the exoplanet to $\tau_0$ at $r=R_0$ by construction (as a boundary condition).

It follows that
\begin{equation}
\begin{split}
A\left(0,R_0\right) &= \pi R_0^2 + \frac{2\pi R_0^2}{\tau_c^2} e^{-\tau_0} \left( 1 - \tau_c - e^{-\tau_c} \right) \\
&\simeq \pi R_0^2 -  \frac{2\pi R_0^2}{\tau_c} e^{-\tau_0}.
\end{split}
\end{equation}
A more illuminating way to write the preceding equation is
\begin{equation}
\frac{A\left(0,R_0\right)}{\pi R_0^2} \simeq 1 - \frac{2 l_{\rm mfp}}{R_0} e^{-\tau_0}.
\end{equation}
The correction terms are small if $\tau_c \ggg 1$ or $l_{\rm mfp} \lll R_0$.  Thus, $A(0,R_0) \simeq \pi R_0^2$ for any value of $\tau_0$.

\subsection{Polytrope of index 1}
\label{subsect:polytrope1}

\begin{figure}
\begin{center}
\vspace{-0.1in}
\includegraphics[width=\columnwidth]{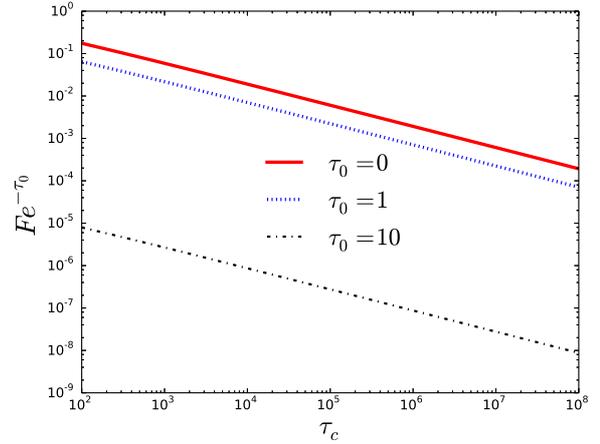}
\end{center}
\vspace{-0.1in}
\caption{Correction factor to the projected area of the spherical exoplanet at $r=R_0$ for a polytrope of index 1 (see text for definition of $F$) as a function of the optical depth difference between the exoplanet center and reference transit radius.  The assumption of $R_0/R=1$ has been made; other choices (e.g., $R_0/R=0.95$) yield similar outcomes.  The $\tau_0=0$ curve isolates the effect of $F$.}
\vspace{-0.1in}
\label{fig:polytrope}
\end{figure}

The mass density profile is (Chapter 4 of \citealt{chandra})
\begin{equation}
\rho = \frac{\rho_c \sin{x}}{x},
\end{equation}
where $\rho_c$ is the mass density at $r=0$, $x \equiv \pi r / R$ and $R$ is the radius of the exoplanet.  By construction, $\rho=0$ when $r=R$.  The corresponding pressure profile is (Chapter 4 of \citealt{chandra})
\begin{equation}
P = \frac{W_1 G M^2}{R^4} \left( \frac{\rho}{\rho_c} \right)^2,
\end{equation}
where $G$ is the gravitational constant and $M$ is the mass of the exoplanet.  The constant $W_1 = 0.392699$ is taken from Table 4 (page 96) of Chapter 4 of \cite{chandra}.  When $r=R_0$ and $x=x_0 \equiv \pi R_0/R$, the reference mass density and pressure are
\begin{equation}
\rho_0 = \frac{\rho_c \sin{x_0}}{x_0}, ~P_0 = \frac{W_1 G M^2}{R^4} \left( \frac{\sin{x_0}}{x_0} \right)^2.
\end{equation}
Since the profiles of mass density and pressure need to join smoothly to the ideal gas law at $r=R_0$, one may solve for the temperature at the reference transit radius,
\begin{equation}
T_0 = \frac{W_1 G M^2}{{\cal R} \rho_c R^4} \left( \frac{\sin{x_0}}{x_0} \right),
\end{equation}
where ${\cal R}$ is the specific gas constant.  This exercise demonstrates that if the interior structure of an exoplanet is a priori known, then the conditions at the reference transit radius are completely specified.

By again imposing the boundary condition that $\tau = \tau_0$ at $r=R_0$, one obtains
\begin{equation}
\tau = \tau_c \left( 1 -  \frac{S}{S_0} \right) + \tau_0,
\label{eq:tau_poly1}
\end{equation}
where the trigonometric integral is
\begin{equation}
S \equiv \int^x_0  \frac{\sin{x^\prime}}{x^\prime} ~dx^\prime.
\end{equation}
The optical depth between the center of the exoplanet and the reference transit radius is $\tau_c \equiv \rho_c \kappa R S_0 / \pi$.  The quantity,  
\begin{equation}
S_0 \equiv S\left(x_0\right),
\end{equation}
depends on the chosen value of $R_0/R$.  Similar to a polytrope of index 0, demanding that $\tau_c \ggg 1$ is the same as assuming
\begin{equation}
l_{\rm mfp} \lll \frac{R_0 S_0}{\pi},
\end{equation}
where $l_{\rm mfp} = 1/\rho_c \kappa$ is the photon mean free path at the \textit{center} of the exoplanet and is thus expected to be very small.  No assumption is made on $\tau_0$.  At $r=0$, the optical depth is again $\tau_c + \tau_0$ and $\tau_0$ serves as its ``zero point" as before.

It follows that
\begin{equation}
\frac{A\left(0,R_0\right)}{\pi R_0^2} = 1 - F e^{-\tau_0},
\end{equation}
where the correction factor involves the integral,
\begin{equation}
F \equiv \frac{2}{\pi^2} \left( \frac{R_0}{R} \right)^2 \int^{\pi R_0/R}_0 x e^{\tau_c \left( S/S_0 - 1 \right)} ~dx.
\end{equation}
Since we expect $R_0/R \sim 1$, it suffices to numerically evaluate $F$ for $R_0/R=1$ as a function of $\tau_c$ (Figure \ref{fig:polytrope}).  When $R_0/R=1$, one obtains $S_0 \approx 1.852$ but in order to evaluate the integral accurately $S_0$ needs to be numerically computed to machine precision.  It is important to note that the correction to $A(0,R_0)/\pi R_0^2=1$ is $F e^{-\tau_0}$.  With $\tau_0=1$, the correction is about $0.37 F$; choices of $\tau_0 \sim 1$--10 will make the correction even smaller (Figure \ref{fig:polytrope}).

Figure \ref{fig:polytrope} shows that the corrections to $A(0,R_0)/\pi R_0^2=1$ become $\lesssim 1\%$ for $\tau_0=1$ when $\tau_c \gtrsim 10^4$.  It is worth estimating conservative values for $\tau_c$,
\begin{equation}
\tau_c \sim 10^8 \left( \frac{\rho_c}{1 \mbox{ g cm}^{-3}} \frac{\kappa}{0.05 \mbox{ cm}^2 \mbox{ g}^{-1}} \frac{P}{1 \mbox{ bar}} \frac{R_0}{R_{\rm J}} \right),
\end{equation}
where $R_{\rm J}=7.1492 \times 10^9$ cm is the radius of Jupiter.  The order-of-magnitude estimate of the opacity, as well as its linear dependence on pressure, is taken from \cite{g05} and is broadly consistent with the more detailed calculations of \cite{valencia13} and \cite{freedman14} for $P =1$ bar and $T \sim 1000$ K.  The actual values for $\rho_c$ and the opacity are likely to be higher. 

\begin{figure}
\begin{center}
\vspace{-0.1in}
\includegraphics[width=\columnwidth]{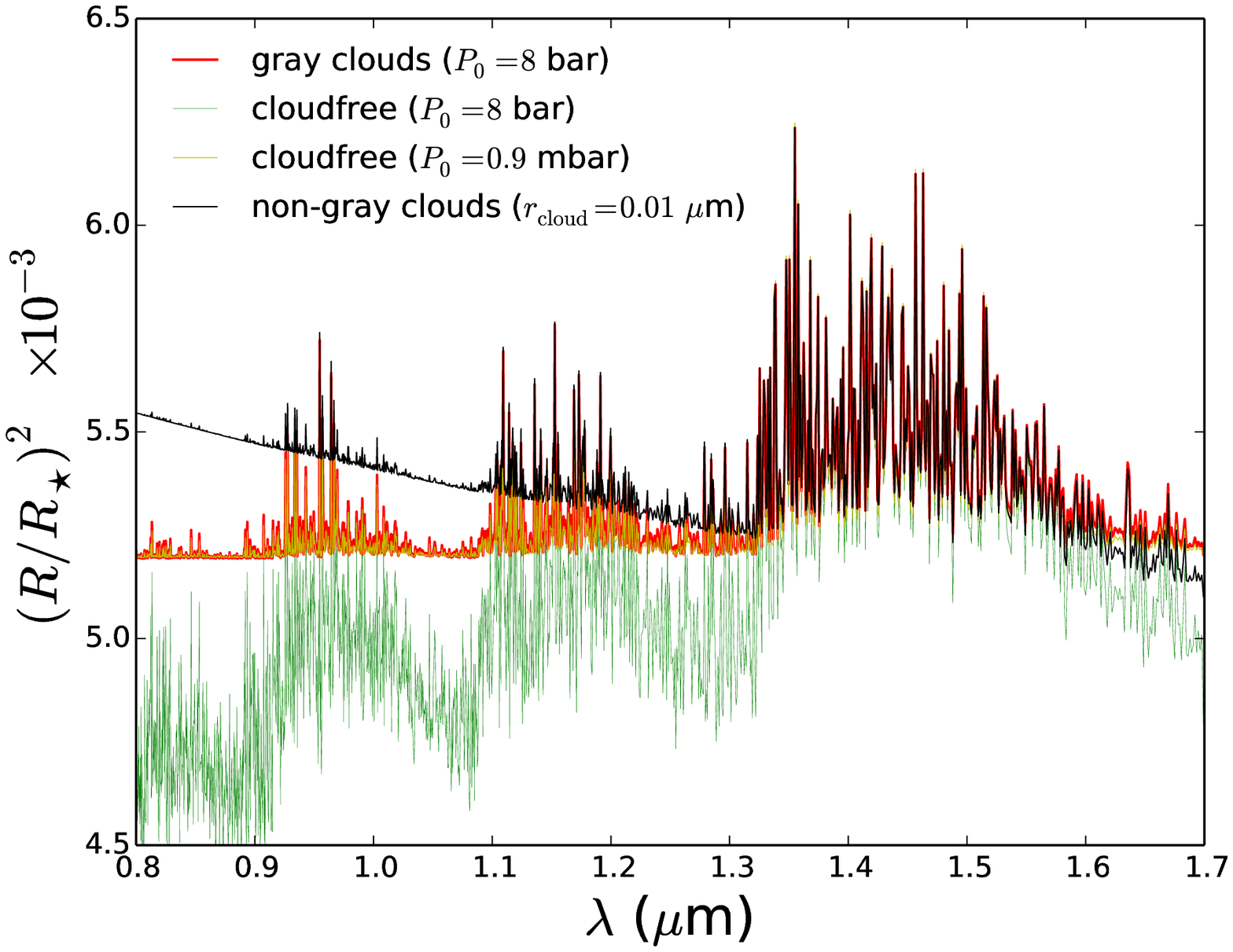}
\includegraphics[width=\columnwidth]{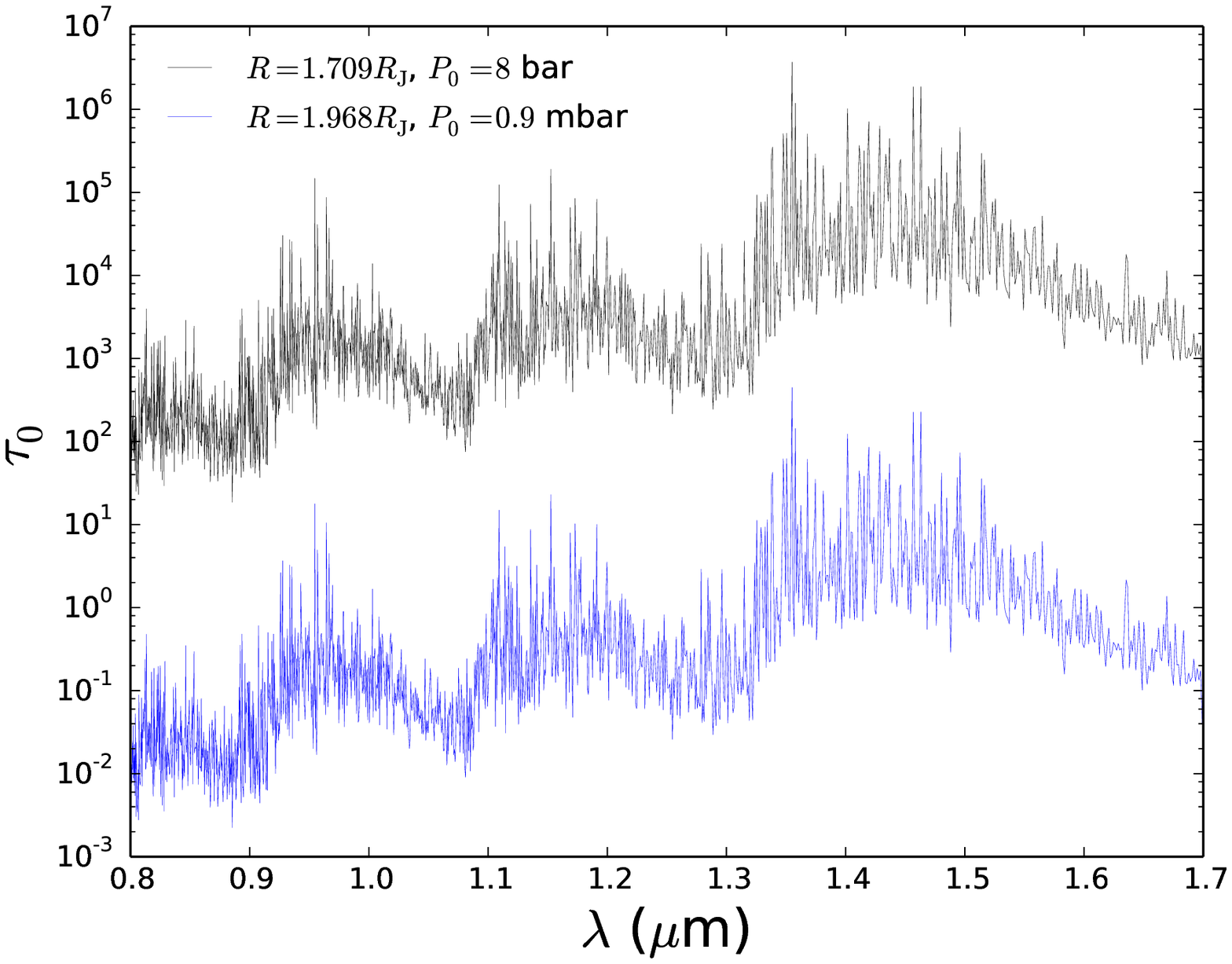}
\end{center}
\vspace{-0.1in}
\caption{Top panel: synthetic transmission spectra adopting parameter values from WASP-17b (see text). Two of these spectra assume $R_0 = 1.709 R_{\rm J}$ and $P_0=8$ bar; one of them is cloudfree, while the other assumes a grey cloud with a constant opacity of 0.01 cm$^2$ g$^{-1}$ corresponding to a transit chord located at 0.9 mbar. The third spectrum uses $R_0 = 1.968 R_{\rm J}$ and $P_0=0.9$ mbar to mimic a transit chord with grey clouds. The fourth spectrum includes non-grey clouds via Mie theory (see text).  Bottom panel: corresponding profiles of $\tau_0$ involving only the water opacity.}
\vspace{-0.1in}
\label{fig:spectra}
\end{figure}

Figure \ref{fig:spectra} shows four examples of hot Jovian transmission spectra, where the absolute transit depths are typically $\sim 10^{-3}$; also shown are two choices for $\tau_0$, which is generally a function of wavelength.  The spectral features are typically $\sim 10^{-4}$ variations in the relative transit depth corresponding to $\sim 10\%$ variations in $\pi R^2$.  A desired property is for the correction to $A(0,R_0)/\pi R_0^2=1$ to be much smaller than these variations in $\pi R^2$.  Figure \ref{fig:polytrope} shows that with $\tau_c=10^8$, one already has $F < 10^{-3}$.  Therefore, $Fe^{-\tau_0} \ll 1$ independent of the value of $\tau_0$ and the correction to $A(0,R_0)/\pi R_0^2=1$ is negligible in the sense that it is much smaller than the variations in the relative transit depth associated with spectral features.

The optical depth increases smoothly from a value of $\tau_c$ at the center of the exoplanet to its boundary-condition value of $\tau_0$ at the reference transit radius.  It is neither constant within $0 \le r \le R_0$ nor discontinuous at $r=R_0$.

\section{Transit radius formula with different expressions for gravity}

Equation (\ref{eq:main}) is derived by solving for $h = A(R_0,\infty)/2 \pi R_0$ and inserting it into $R = R_0 + h$.  Evaluating
\begin{equation}
A\left(R_0,\infty\right) = \int^{\infty}_{R_0} \left( 1 - e^{-\tau} \right) 2 \pi r ~dr 
\end{equation}
requires that one elucidates the relationship between $\tau$ and $r$.  Within the atmosphere, assuming the ideal gas law and hydrostatic balance yields
\begin{equation}
\ln{\left( \frac{\tau}{\tau_0} \right)} = - \int^r_{R_0} \frac{mg}{k_{\rm B} T} ~dr,
\end{equation}
where $m$ is the mean molecular mass, $k_{\rm B}$ is the Boltzmann constant and $T$ is the temperature.  Evaluating the integral requires that one specifies $g(r)$.

There are three different ways of expressing the acceleration due to gravity: constant $g$, constant exoplanet mass ($g \propto 1/r^2$) or constant bulk density ($g \propto r$).  Consider a gas-giant exoplanet where $R_0 \sim R_{\rm J}$.  If $P_0 \sim 10$ bar and the infrared photosphere is located at $\sim 1$ mbar, then the atmosphere is $\sim 10$ pressure scale heights thick.  Since $H/R \sim 0.01$ (with $H = k_{\rm B} T / mg$ being the pressure scale height), this means that the atmosphere is $\sim 0.1 R_{\rm J}$ thick.  In the constant exoplanet mass or bulk density approximations, this implies that the acceleration due to gravity is changing by $\sim 10\%$ within the atmosphere, which provides the motivation for investigating these three ways of deriving $\tau(r)$.

In the standard derivation where $g$ is assumed to be constant, one obtains the usual expression for hydrostatic equilibrium,
\begin{equation}
\ln{\left( \frac{\tau}{\tau_0} \right)} = \frac{R_0 - r}{H}.
\label{eq:tau_atmos1}
\end{equation}
One recovers equation (S.4) of \cite{ds13}, equation (20) of \cite{bs17} or equation (8) of \cite{hk17},
\begin{equation}
\begin{split}
h &= H \int^{\tau_0}_0 \frac{1 - e^{-\tau}}{\tau} \left[ 1 + \frac{H}{R_0} \ln{\left( \frac{\tau_0}{\tau} \right)} \right] ~d\tau \\
&\simeq H \left[ \gamma + E_1\left(\tau_0\right) + \ln{\tau_0} \right],
\end{split}
\label{eq:constant_g}
\end{equation}
where the second, approximate equality holds if one assumes the term involving the logarithm in the integrand to be smaller by a factor of $H/R_0$ and is hence dropped, which allows the integral to be evaluated analytically. 

One could instead assume that $g = GM/r^2$ with a constant $M$, which yields
\begin{equation}
\ln{\left( \frac{\tau}{\tau_0} \right)} = \frac{R_0}{H_0} \left( \frac{R_0}{r} -1 \right),
\label{eq:tau_atmos2}
\end{equation}
where $H_0 \equiv k_{\rm B} T / m g_0$, $g_0 \equiv GM/R_0^2$ and $G$ is the gravitational constant.  It follows that
\begin{equation}
\begin{split}
h &= H_0 \int^{\tau_0}_0 \frac{1 - e^{-\tau}}{\tau} \left[ 1 + \frac{H_0}{R_0} \ln{\left( \frac{\tau}{\tau_0} \right)} \right]^{-3} ~d\tau \\
&\simeq H_0 \left[ \gamma + E_1\left(\tau_0\right) + \ln{\tau_0} \right].
\end{split}
\label{eq:constant_mass}
\end{equation}
Again, the integral may only be evaluated analytically if the $\sim H_0/R_0$ term within the integral is dropped.

Alternatively, one may assume the mass of the exoplanet to be given by $M = 4 \pi \bar{\rho} r^3 / 3$, where $\bar{\rho}$ is an average bulk mass density that is assumed to be constant for $r \ge R_0$.  This assumption yields $g = 4 \pi G \bar{\rho} r / 3$, which yields
\begin{equation}
\ln{\left( \frac{\tau}{\tau_0} \right)} = \frac{R_0}{2H_0} \left( 1 - \frac{r^2}{R_0^2} \right).
\label{eq:tau_atmos3}
\end{equation}
It follows that 
\begin{equation}
\begin{split}
h &= H_0 \int^{\tau_0}_0 \frac{1 - e^{-\tau}}{\tau} ~d\tau\\
&= H_0 \left[ \gamma + E_1\left(\tau_0\right) + \ln{\tau_0} \right],
\end{split}
\label{eq:variable_g}
\end{equation}
where we again have $H_0 \equiv k_{\rm B} T / m g_0$, but $g_0 \equiv GM_0/R_0^2$ and $M_0$ is the mass of the exoplanet enclosed by $r=R_0$.  There is no $\sim H/R_0$ or $\sim H_0/R_0$ correction term to drop and the integral is evaluated exactly.

The constant $g$, $g = GM/r^2$ and $g = 4 \pi G \bar{\rho} r / 3$ approaches yield the same result for $h$ and hence $R$, despite the different functional forms for $\tau(r)$.  This is a mathematical coincidence and arises only because small correction terms in the integrand for $A(R_0,\infty)$ were dropped in order to evaluate the integral analytically.  

The optical depth may be constructed using equation (\ref{eq:tau_poly0}) or (\ref{eq:tau_poly1}) for $0 \le r \le R_0$ and equation (\ref{eq:tau_atmos1}), (\ref{eq:tau_atmos2}) or (\ref{eq:tau_atmos3}) for $r \ge R_0$.  For all 6 combinations, the optical depth is continuous across the reference transit radius and finite everywhere.

\section{Treatment of clouds in transmission spectra}

\subsection{Preamble}

From equation (12) of \cite{hk17}, the reference optical depth within equation (\ref{eq:main}) is
\begin{equation}
\tau_0 = \frac{\kappa P_0}{g} \sqrt{\frac{2 \pi R_0}{H}}.
\end{equation}
Note that equation (9) of \cite{ds13} expresses $\tau_0$ (denoted by them as $A_\lambda$) in terms of a reference number density and cross section.  \cite{bs17} write $\tau_0$ as $\tau_s$ in their equation (26), but do not explicitly provide an expression for it beyond their equation (42).  The normalisation degeneracy \citep{bs12,g14,hk17,fh18}, which is the three-way degeneracy between $R_0$, $P_0$ and $\kappa$ (which contains the relative abundances of atoms and molecules) is not explored in detail by either \cite{ds13} or \cite{bs17}.

The wavelength-, temperature- and pressure-dependent opacity function is 
\begin{equation}
\kappa = X_{\rm cloud} \frac{\sigma_{\rm cloud}}{m} + \sum_i \kappa_i X_i \frac{m_i}{m}.
\label{eq:opacity}
\end{equation}
The sum is over all of the atoms, ions and molecules in the atmosphere.  The opacity of each species is denoted by $\kappa_i$.  The volume mixing ratio of each species is $X_i$; it is worth noting that the mass mixing ratio is $X_i m_i / m$, where $m_i$ is the mass of each species.  The cloud volume mixing ratio and cross section are denoted by $X_{\rm cloud}$ and $\sigma_{\rm cloud}$, respectively.  In the current study, the only molecule considered is water as this suffices to construct the necessary arguments.

Assuming a monodisperse cloud (i.e., particles of a single radius), the cloud cross section is 
\begin{equation}
\sigma_{\rm cloud} = Q \pi r_{\rm cloud}^2,
\end{equation}
where $Q$ is the extinction efficiency.  It may be computed using Mie theory (e.g., \citealt{kh18}).  \cite{kh18} provide a convenient fitting function,
\begin{equation}
Q = \frac{Q_1}{Q_0 X^{-a} + X^{0.2}},
\end{equation}
which is calibrated to full numerical calculations.  The dimensionless size parameter is given by $X = 2 \pi r_{\rm cloud} / \lambda$.  This fitting function for $Q$ smoothly connects the regimes of small ($X \ll 1$; Rayleigh) and large ($X \gg 1$) particles.  As an illustration, I adopt the calibration for forsterite (Mg$_2$SiO$_4$): $Q_0 = 11.95$, $Q_1 = 4.16$ and $a=4.05$ (see Table 2 of \citealt{kh18}).

It is worth noting that Mie theory specifies the wavelength dependence of the cloud cross section, but not its spatial dependence.  The latter is driven by poorly known details of the formation, evolution and interaction of the cloud with radiation hydrodynamics and disequilibrium chemistry \citep{marley13,helling18}.  It is possible to prescribe the spatial boundaries of the cloud deck in a phenomenological manner, as has been implemented in the study of the atmospheres of brown dwarfs (e.g., \citealt{burrows06,burrows11}).  Over a limited wavelength range and at low spectral resolution, such as by the Hubble Space Telescope Wide Field Camera 3 (HST-WFC3), it has been shown that the transmission spectrum probes a limited range of pressures and the transit chord may be approximated as being isobaric \citep{hk17}, rendering the spatial dependence of the cloud cross section a non-issue.

\subsection{Transmission spectra}

For clarity of discussion, the specific case study of WASP-17b is used.  \cite{fh18} have previously estimated that $R_0 = 1.709 R_{\rm J}$ at $P_0=8$ bar based on the inference made by \cite{heng16} that the HST-STIS transit chord of WASP-17b is cloud-free (see also \citealt{fh19}).  The surface gravity of WASP-17b is $g=316$ cm s$^{-2}$, while the stellar radius of WASP-17 is $R_\star=1.583 ~R_\odot$ \citep{southworth12}.   A temperature of $T=1700$ K is adopted, which is roughly the retrieved transit chord temperature reported by \cite{fh18} for WASP-17b (see their Table 2).  For illustration, I adopt $X_{\rm H_2O}=10^{-3}$, which is not an uncommon value for the retrieved volume mixing ratio of water (see Figure 29 of \citealt{fh18}).  With these numbers, $H \approx 2000$ km and $H/R_0 \approx 0.02$.

Figure \ref{fig:spectra} shows a pair of transmission spectra with $R_0 = 1.709 R_{\rm J}$ and $P_0=8$ bar.  One of these spectra adds a constant term of $\kappa_{\rm cloud}=0.01$ cm$^2$ g$^{-1}$ to the opacity function in equation (\ref{eq:opacity}), which represents a grey cloud comprising large particles.  For a grey cloud, the cloud cross section and mixing ratio may be subsumed into a single number.  Even though this grey cloud is assumed to be spatially uniform, it corresponds to a pressure of 0.9 mbar for the transit chord using equation (\ref{eq:p_chord}).  One may \textit{mimic} the effect of this grey cloud by setting $P_0=0.9$ mbar and using hydrostatic balance to set the correct value of $R_0$ that corresponds to this pressure ($R_0 = 1.968 R_{\rm J}$), as shown in Figure \ref{fig:spectra}.  It is important to note that it is the $E_1(\tau_0)$ term in equation (\ref{eq:main}) that allows for this mimicry.

There are two concerns with this mimicry.  First, the value of $\kappa_{\rm cloud}$ may be specified from first principles by specifying the cloud particle radius and computing the extinction efficiency and cross section using Mie theory, whereas it is less clear how intrinsic cloud properties may be related to $R_0$.  Second, the mimicry does not extend to clouds comprising small particles.  In Figure \ref{fig:spectra}, an example is shown with $r_{\rm cloud}=0.01$ $\mu$m and $X_{\rm cloud} = 10^{-16}$.  This cloud produces a non-flat spectral continuum between 0.8 and 1.3 $\mu$m.  It is difficult to see how such wavelength-dependent behaviour may be specified from first principles via $R_0$.

The reference transit radius and reference pressure are not independent quantities.  Rather, they specify a wavelength-independent reference point within the exoplanet, analogous to the radiative-convective boundary in gas giants.  From a phenomenological point of view, the role of a wavelength-independent $R_0(P_0)$ is well-established in atmospheric retrievals \citep{fh18}. If one needs to fit for a different value of $R_0$ at each wavelength, then the number of fitting parameters will always exceed the number of data points.

\section{Implications}

There are several implications of the current study.
\begin{enumerate}

\item The second footnote of \cite{je18} may be disregarded.

\item Equation (\ref{eq:main}) may be used without assuming $\tau_0 \gg 1$ and the $E_1(\tau_0)$ term may be retained.

\item The $E_1(\tau_0)$ term in equation (\ref{eq:main}) should not be used as a proxy for a cloud deck.

\item Atmospheric retrievals should continue to fix the value of $R_0$ or $P_0$ and include the other quantity as a fitting parameter \citep{fh18}, unless the interior structure of the exoplanet is a priori known.

\end{enumerate}

\vspace{0.1in}
\noindent
\textit{The anonymous referee is credited with providing three thorough reviews that led to an improved manuscript.  Besides the intellectual stimulation provided by \cite{je18}, I am grateful to Andres Jord\'{a}n and N\'{e}stor Espinoza for useful and constructive conversations.  I benefited from a colloquium given on 28th February 2019 at the Jet Propulsion Laboratory, where Yan B\'{e}tr\'{e}mieux was part of the audience.  I thank Yann Alibert for useful email exchanges on polytropes and the Lane-Emden equation.  I acknowledge partial financial support from the Center for Space and Habitability (CSH), the PlanetS National Center of Competence in Research (NCCR), the Swiss National Science Foundation, the MERAC Foundation and an European Research Council (ERC) Consolidator Grant (number 771620).}

\label{lastpage}
\end{document}